\documentclass[11pt,dvips]{article}

\usepackage{epsfig,times} 
\usepackage{picinpar}
\usepackage{wrapfig}
\usepackage{floatflt}
\makeatletter
\renewcommand{\section}{\@startsection{section}{1}{0in}
	{0.4\baselineskip}{0.1\baselineskip}{\Large\bf}}
\renewcommand{\subsection}{\@startsection{subsection}{2}{0in}
	{0.25\baselineskip}{-\baselineskip}{\large\bf}}
\renewcommand{\subsubsection}{\@startsection{subsubsection}{3}{0in}
	{0.1\baselineskip}{-\baselineskip}{\normalsize\bf}}
\makeatother
\begin{document}

%
\makeatletter\newcommand{\ps@icrc}{
\renewcommand{\@oddhead}{\slshape{ICRC 1999 Evening Workshop Session talk
(updated version)}\hfil}}
\makeatother\thispagestyle{icrc}
%
%

\begin{center}
%
{\LARGE \bf Deformed Lorentz Symmetry\\ 
and High-Energy Astrophysics (I)}
\end{center}

\begin{center}
%
%
{\bf L. Gonzalez-Mestres$^{1,2}$}\\
{\it $^{1}$Laboratoire de Physique Corpusculaire, Coll\`ege de France, 75231 Paris Cedex 05, France\\
$^{2}$L.A.P.P., B.P. 110, 74941 Annecy-le-Vieux Cedex, France}
\end{center}

\begin{center}
{\large \bf Abstract\\}
\end{center}
\vspace{-0.5ex}
%
%
Lorentz symmetry violation (LSV) 
can be generated at the Planck scale, or at some
other fundamental length scale, and nevertheless
naturally preserve Lorentz symmetry
as a low-energy limit (deformed Lorentz symmetry, DLS). Deformed
relativistic kinematics (DRK) would then be consistent with special relativity
in the limit $k$ (wave vector) $\rightarrow ~0$ and allow for a deformed
version of general relativity and gravitation. 
If LSV is a very high-energy, very low-distance phenomenon, it is expected 
to be driven by energy-dependent parameters and, if at all detectable, to
produce the cleanest signatures at the highest attainable energies. 
We present an updated discussion of the
possible implications of this pattern for high-energy cosmic-ray physics,
focusing mainly on
an approach where the LSV parameter varies like the square 
of the energy scale (quadratically deformed relativistic kinematics, 
QDRK).
It turns out that a $\approx ~10^{-6}$ LSV at Planck scale, leading to
a DLS pattern, would potentially be
enough to produce very important observable effects on the properties of
cosmic rays at the $\approx ~10^{20}~eV$ scale (absence of GZK cutoff,
stability of unstable particles, lower interaction rates, kinematical
failure
of any parton model as well as of standard formulae for Lorentz contraction
and time dilation...). Possible neutrino pulses at energies up to
$\approx ~10^{19}~eV$ 
from $\approx ~10^{20}~eV$
protons accelerated by gamma-ray bursts, 
with neutrino arrival times implying energy-dependent delays
from DRK, are also discussed. 
Although ultra-high energy cosmic rays (UHECR) appear 
to be the most appropriate probe to test Planck-scale LSV, we also discuss 
suggestions to explore it at the observed gamma-ray burst (GRB)
photon energies using models with 
a LSV parameter proportional to the energy 
scale (linearly deformed relativistic kinematics, LDRK).   

~

This paper updates and further develops contributions H.E. 1.3.16 and OG
3.1.10 published in the Proceedings of the ICRC 1999 Conference, Salt-Lake
City August 1999 
(Gonzalez-Mestres, 1999a and 1999b).
~
~
\vspace{1ex}
~
%
%
\section{What is relativity?}
\label{relativity.sec}

~

Arguments used in the priority debate: "who was (were)
the author(s) of the special relativity theory?" tacitly involve, without
really addressing it, a fundamental physics issue. It is implicitly assumed
that the basic physics behind special relativity is exactly what standard 
textbooks have been teaching in the last eight decades: Lorentz symmetry 
would be an abstract, 
intrinsic property of space-time that matter cannot escape.
In this approach, all particles are compelled to move inside the minkowskian 
space-time.
However, such a non-trivial interpretation of special relativity
is not a well-established physical law and there is no proof
of its absolute validity. 
It does not correspond to the initial formulation of the Poincar\'e
relativity principle (1895-1905), 
and
current particle theory suggests that
Lorentz symmetry can be violated.
A close look reveals that 
historical arguments are biased by physical prejudices and arbitrary
interpretations.  

The French mathematician Henri Poincar\'e was the first author to consistently
formulate the relativity principle, stating
(Poincar\'e, 1895):
{\it
"Absolute motion of matter, or, to be more precise, the
relative motion of weighable matter and ether, cannot be disclosed. All that
can be done is to reveal the motion of weighable matter with respect to
weighable matter"}. Such a revolutionary claim was not easily accepted. 
Poincar\'e was fighting for a decade to convince 
other scientists as well as the public opinion.
He further emphasized the deep content of this new and, at the time,  
unconventional law of Nature 
when he wrote (Poincar\'e, 1901):
{\it "This principle will be confirmed with increasing precision,
as measurements
become more and more accurate"}.

Although textbooks and press usually present special relativity as having been
formulated in the celebrated Einstein's 1905 paper, several authors have 
emphasized the actual role of H. Poincar\'e 
in building relativity theory previous to Einstein
and the relevance of Poincar\'e's thought
(Logunov, 1995 and 1997; Feynmann, Leighton, \& Sands,    
1964).
In his
June 1905 paper (Poincar\'e, 1905), published before Einsteins's article
(Einstein, 1905) arrived (on June 30) to the editor,
Henri Poincar\'e explicitly wrote the relativistic transformation law for the
charge density and velocity of motion and applied to gravity
the "Lorentz group", assumed to hold for "forces of whatever origin". 
All the ingredients of special relativity, as well as its basic original 
concepts,
are clearly formulated in this work which, furthermore, emphasizes 
the need of 
a new, relativistic, theory of gravitation.
But Poincar\'e's priority is sometimes
denied on the scientific grounds that {\it "Einstein essentially
announced the failure of all ether-drift experiments past and future as a
foregone conclusion, contrary to Poincar\'e's empirical bias"} (Miller, 1996),
that Poincar\'e did never {\it "disavow the ether"} (Miller, 1996) or that
{\it "Poincar\'e never challenges... the absolute time of newtonian
mechanics... the ether is not only the absolute space of mechanics... but a
dynamical entity"} (Paty, 1996). Is this argumentation correct, is it 
based on 
well-established physical evidence? We do not think so. 
In fact, these authors implicitly assume that A. Einstein
was right in 1905 when {\it "reducing ether to the absolute space of
mechanics"} (Paty, 1996) and that H. Poincar\'e was wrong because {\it "the
ether fits quite nicely into Poincar\'e's view of physical reality: the ether
is real..."} (Miller, 1996). But, with the present status of particle physics
and cosmology, as well as of condensed-matter physics and of the theory 
of dynamical systems, there is no  
scientific evidence for Einstein's 1905 absolute,
"geometric" view of relativity. 
The existence of a physical 
"ether", playing an important dynamical role, would not be 
incompatible with the
existing (low-energy) experimental evidence for the relativity principle.

~

\subsection{Particle physics point of view:}
\label{pp.sec}

~

Modern particle physics has
brought back the concept of a non-empty vacuum where free particles propagate:
without such an "ether" where fields can condense, the standard model
of electroweak interactions could not be written and quark confinement could
not be understood.  
The mechanism producing
the masses of the W and Z bosons is close to 
the Meissner effect, where the condensed Cooper pairs 
(equivalent to the Higgs field)
prevent the magnetic 
field (virtual photons)
from propagationg beyond a certain distance (the London length)
inside a superconductor: in the case of the standard model, the effect is
mainly observed through the masses (inverse London lengths)
of the intermediate bosons propagating in the (vacuum) Higgs field condensate.
Furthermore, modern cosmology is not incompatible with the idea of
an "absolute local frame" close to that suggested by the study of cosmic
microwave background radiation (see f.i. Peebbles, 1993). 

Therefore, the "ether" may well turn out
to be a real entity in the {\it XXI-th} century physics and astrophysics. 
Then, the relativity principle would become
a symmetry of physics, another revolutionary concept whose paternity was 
attributed to
H. Poincar\'e by R.P. Feynman (as quoted by Logunov, 1995):
{\it "Precisely Poincar\'e proposed investigating what could be done with the
equations without altering their form. It was precisely his idea to pay
attention to the symmetry properties of the laws of physics"}. Actually, 
Poincar\'e did even more: in all his papers since 1895, 
he emphasized another deep concept: the {\it dynamical}
origin of special relativity. 

Dynamics is, by definition, a scale-dependent
property of matter. In a global view of physics, a dynamical property 
of matter would be the opposite concept to an intrinsic, 
geometric property of space-time.
A basic, unanswered question for particle physics is therefore:
was Poincar\'e right when, in his papers 
since 1895 and in particular in his
note of June 1905 strongly premonitory of nowadays
grand-unified theories, he considered the relativity
principle as a {\it dynamical} phenomenon, related to a common origin of all 
the
existing forces?

As symmetries in particle physics are in general violated,
Lorentz symmetry may be broken and an absolute local rest frame may
be detectable through experiments performed beyond some critical scale
or close to that scale. 
Poincar\'e's special relativity (a symmetry applying to
physical processes) could live with this situation, but not Einstein's
approach such as it was formulated in 1905 (an absolute geometry of space-time
that matter cannot escape). But, how to check whether Lorentz symmetry is
actually broken?
We discuss here two issues: a) the scale
where we may expect Lorentz symmetry to be violated and the scale(s)
at which the effect may be observable; b) the physical phenomena
and experiments potentially able to uncover Lorentz symmetry violation (LSV).
Previous papers on the subject are (Gonzalez-Mestres, 1998a, 1998b, 1998c
and 1999a) 
and references therein. We have proposed that Lorentz symmetry be a low-energy
limit, broken following a $k^2$-law ($k$ = wave vector) between the low-energy
region and some fundamental energy (length) scale.  

~

\subsection{Condensed-matter point of view:}
\label{cm.sec}

~

It seems obviously justified to examine what could be a "condensed-matter point 
of view", as particle physics and cosmology have used many condensed-matter
analogues in the last five decades (e.g. the pattern of spontaneous symmetry
breaking) and the Klein-Gordon equation is a typical equation for 
wave
propagation. Particle physics uses nowadays concepts such as strings, vortices,
monopoles, topological defects... which are closely related to condensed-matter 
phenomena and make reference to the 
"vacuum" as a material medium.

Lorentz symmetry, viewed as a property of dynamics,
implies no reference to absolute
properties of space and time
(Gonzalez-Mestres, 1995). In a two-dimensional
galilean space-time,
the wave equation:
\equation
\alpha ~\partial ^2\phi /\partial t^2~-~\partial ^2\phi /\partial x^2 = F(\phi )
\endequation
\noindent
with $\alpha$ = $1/c_o^2$ and $c_o$ = critical
speed, remains unchanged under "Lorentz" transformations leaving
invariant the squared
interval
$ds^2 = dx^2 - c_o^2 dt^2$ .
Any form of matter made with solutions of equation (1) , built in 
the laboratory in a set-up at rest, 
would feel a relativistic space-time even if the real space-time is
galilean and if an absolute rest frame exists in the
underlying dynamics beyond the wave equation.
The solitons of the sine-Gordon equation are
obtained taking in (1) :
\equation
F(\phi )~~ = ~~-~(\omega /c_o)^2~sin~\phi
\endequation
\noindent
$\omega $ being a characteristic frequency of the dynamical system.
The two-dimensional universe made of such sine-Gordon solitons
would indeed behave like a two-dimensional minkowskian
world with the laws of special relativity.
The actual structure of space and time
can only be found by going
to deeper levels of resolution where the equation fails,
similar to the way
high-energy accelerator experiments explore the inner structure of
"elementary" particles (but cosmic rays have the highest attainable
energies). As modern particle physics views "elementary" particles
as excitations of vacuum, there would be no inconsistency in assumig that
the space-time felt by such particles is similar to the soliton analogy.
and does not have and absolute meaning.
In such a scenario, that cannot be ruled out by any present experiment,
superluminal sectors of matter can exist and even be its ultimate
building blocks. This clearly makes sense, as:
a) in a
perfectly transparent crystal, at least two critical speeds can be
identified, those of light and
sound; b) the potential approach to
lattice dynamics in solid-sate physics
is precisely the form of electromagnetism
in the limit $c_s~c^{-1}~\rightarrow ~0$ , where $c_s$ is the speed of sound
and $c$ that of light. See, for instance, (Gonzalez-Mestres, 1999a) and
references therein.

At the fundamental lenght scale, and taking a simplified
two-dimensional illustration, gravitation may even be a composite
phenomenon (Gonzalez-Mestres, 1997a), related for instance to
fluctuations of the parameters of equations like:
\equation
A~d^2/dt^2~~[\phi ~(n)]~+~H~d/dt~[~\phi ~(n+1)~-~\phi ~(n-1)]~-~\Phi_n ~[\phi ]~=~
0
\endequation
where we have quantized space to schematically account
for the existence of the fundamental length $a$, $\phi$ is a wave function,
$n$ designs by an integer lattice sites spaced by a distance
$a$, $A$ and $H$ are
coeficients and
$\Phi_n ~[\phi ]$ is defined by:
\equation
\Phi_n ~[\phi ]~=~K_{fl}~[2~\phi ~(n)~-~\phi
~(n-1)~-~\phi ~(n+1)]~+~\omega _{rest}^2~\phi
\endequation
$K_{fl}$ being a coefficient and $(2\pi )^{-1}~\omega _{rest}$ a rest frequency.

In the continuum limit, the coefficients $A~=~g_{00}$ , $H~=~g_{01}~
=~g_{10}$ and $-K_{fl}~=~g_{11}$ can be regarded as the matrix elements of
a space-time bilinear metric with equilibrium values: $A~=~1$ , $H~=~0$ and
$K_{fl}~=~K$~. Then, a small local fluctuation:
\equation
A~=~1~+~\gamma
\endequation
\equation
K_{fl}~=~K~(1~-~\gamma )
\endequation
with $\gamma ~\ll ~1$ would be equivalent to a small, static
gravitational field
created by a far away source. 

In conclusion, our present knowledge of condensed-matter physics does not plead
in favour of Einstein's 1905 , intrinsically geometric, approach to relativity.
It instead suggests that
Poincar\'e was right in not ruling out the Ether and in developing instead the
concept of physical symmetry (the "Lorentz group").    

~

\section{Lorentz Symmetry As a Low-Energy Limit}
\label{deformed.sec}

~

Low-energy tests of special relativity have confirmed its validity to an
extremely good accuracy ($\approx ~10^{-21}$ from nuclear magnetic
resonance experiments), but the situation at very high energy remains 
unclear. Not only high-energy measurements
are less precise, but the violation of the
relativity principle can be driven by energy-dependent parameters.
To discuss possible Lorentz symmetry violation, the hypothesis
of a preferred refrence 
frame seems necessary. In what follows, all discussions 
are performed in this frame, that we assume to be close to the natural
cosmological one defined by cosmic background radiation 
(see, f.i. Peebles 1993).

If Lorentz symmetry violation (LSV) follows
a $E^2$ law
($E$ = energy), similar to the effective gravitational coupling, it can
be $\approx ~1$ at $E~\approx ~10^{21}~eV$ (just above the highest
observed cosmic-ray energies) and $\approx ~10^{-26}$ at $E~
\approx ~100~MeV$ (corresponding to the highest momentum scale involved in
nuclear magnetic resonance experiments). Such a pattern of LSV 
(deformed Lorentz symmetry, DLS) will escape all
existing low-energy bounds. 
If LSV is of order 1
at Planck scale ($E~
\approx ~10^{28}~eV$), and following a similar law, it will be $\approx
~10^{-40}$ at $E~\approx ~100~MeV$ . Our suggestion is not in contradiction
with Einstein's thought such as it became after he had developed general
relativity. In 1921 , A. Einstein wrote in "Geometry and Experiment" 
(Einstein, 1921):
{\it "The interpretation of geometry advocated here cannot be directly applied
to submolecular spaces... it might turn out that such an extrapolation is
just as incorrect as an extension of the concept of temperature to particles
of a solid of molecular dimensions"}. The absoluteness of the minkowskian
space-time was clearly abandoned through this statement.
It is in itself remarkable that special relativity holds at the
attained accelerator energies, but there is no fundamental reason for this to
be the case above Planck scale.

~
\subsection{Deformed relativistic kinematics}
\label{drk.sec}

~

A typical example of patterns violating Lorentz symmetry at very short distance
is provided by nonlocal models where an absolute local rest frame exists and
non-locality in space is introduced through a
fundamental length scale $a$ where new physics 
is expected to occur (Gonzalez-Mestres, 1997a). 
Such models naturally lead to a deformed
relativistic kinematics (DRK) of the form (Gonzalez-Mestres, 1997a and 1997b):
\equation
E~=~~(2\pi )^{-1}~h~c~a^{-1}~e~(k~a)
\endequation
\noindent
where $h$ is the Planck constant, $c$ the speed of light, $k$ the wave vector,
and
$[e~(k~a)]^2$ is a convex
function of $(k~a)^2$ obtained from vacuum dynamics.
Such an expression is equivalent to special relativity in the
small $k$ limit. 
Expanding equation (1) for $k~a~\ll ~1$ , we can write 
(Gonzalez-Mestres, 1997a and 1997c):
\begin{eqnarray}
e~(k~a) & \simeq & [(k~a)^2~-~\alpha ~(k~a)^4~ 
+~(2\pi ~a)^2~h^{-2}~m^2~c^2]^{1/2}
\end{eqnarray}
\noindent
$\alpha $ being a model-dependent constant, in the range $0.1~-~0.01$ for
full-strength violation of Lorentz symmetry at the fundamental length scale,
and {\it m} the mass of the particle. For momentum $p~\gg ~mc$ , we get:
\begin{eqnarray}
E & \simeq & p~c~+~m^2~c^3~(2~p)^{-1}~ 
-~p~c~\alpha ~(k~a)^2/2~~~~~
\end{eqnarray}
The "deformation" approximated by 
$\Delta ~E~=~-~p~c~\alpha ~(k~a)^2/2$ in the right-hand
side of (9) implies a Lorentz symmetry violation in the ratio $E~p^{-1}$
varying like $\Gamma ~(k)~\simeq ~\Gamma _0~k^2$ where $\Gamma _0~
~=~-~\alpha ~a^2/2$ . If $c$ is a universal parameter for all
particles, the DRK defined by (7) - (9) preserves Lorentz symmetry 
in the limit $k~\rightarrow ~0$, contrary to the standard
$TH\epsilon \mu $ model (Will, 1993). If, besides $c$ , 
$\alpha $ is also universal,
LSV does not lead (Gonzalez-Mestres, 1997a, c and e)
to the spontaneous decays predicted in
(Coleman, \& Glashow, 1997 and subsequent papers) at ultra-high energy
using a $TH\epsilon \mu $-type approach. 
On more general grounds,
as we also pointed out, the existence of very high-energy cosmic rays
can by no means 
be regarded as an evidence against LSV, as the relevant kinematical
balances can be sensitive to many small parameters. In particular,
any form of LSV should be considered (Gonzalez-Mestres, 1997d and 1997e)
and not only (like in the papers by Coleman and Glashow)
models driven by low-energy 
constant parameters. 

As previously emphasized, 
the above non-locality may actually be an approximation to
an underlying dynamics involving superluminal particles
(Gonzalez-Mestres, 1996, 1997b, 1997f and 1997g), 
just as electromagnetism looks nonlocal
in the potential approximation to lattice dynamics in solid-state physics:
it would then correspond to the limit $c~c_i^{-1}~\rightarrow~0$
where $c_i$ is the superluminal critical speed. Contrary to the
$TH\epsilon \mu $-type scenario considered
by Coleman and Glashow, where LSV
occurs explicitly in the hamiltonian already at $k~=~0$ through
a non-universality of the critical speed in vacuum 
(the rest masses of charged particles are 
no longer given by the relation $E~=~m~c^2$ , $c$ being the speed
of light in the $k~\rightarrow 0$ limit), our DLS
approach can preserve standard gravitation and general relativity as
low-energy limits. 
Furthermore, phenomenological estimates
by Coleman and Glashow do not consider 
possible deformations of the relativistic kinematics: they use 
undeformed kinematics for single particles at ultra-high
cosmic-ray energies. More recent   
(1998) papers by these authors bring no new result as compared to our
1997 papers and present the same fundamental limitation as their 1997 article. 
Physically, the two approaches are really different:
by choosing a $TH\epsilon \mu $ scenario, Coleman and Glashow implicitly
assume that Lorentz symmetry is broken in an "external" way,
by a small {\it macroscopic } 
effect. On the contrary, as explained above, our pattern 
attributes LSV to an "internal" very high-energy, very low-distance  
phenomenon completely disappearing at macroscopic scale.    
 
A fundamental question is whether $c$ and $\alpha $ are universal. This may be
the case for all "elementary" particles, i.e.
quarks, leptons, gauge bosons...,
but the situation is less obvious for hadrons, nuclei and heavier objects.
From a naive soliton model (Gonzalez-Mestres, 1997b and 1997f), 
we inferred that: a) $c$ is
expected to be universal up to extremely 
small corrections ($\sim 10^{-40}$ , far 
below the values considered by Coleman and Glashow)
escaping all existing bounds; b)
an
approximate rule can be to take $\alpha $ universal for leptons, gauge bosons
and light hadrons (pions, nucleons...) and assume a $\alpha \propto m^{-2}$
law for nuclei and heavier objects, the nucleon mass setting the scale.
With this rule, DRK introduces no anomaly in the relation between inertial
and gravitational masses at large scale (Gonzalez-Mestres, 1998c). Basically,
the $\alpha \propto m^{-2}$ law makes compatible DRK for a large body with
a similar DRK for smaller parts of it which, otherwise, could not travel at
the same speed as the whole body if the relation 
$v~=~dE/dp$ (Gonzalez-Mestres, 1997a , 1997d and 1997f) is used. 

The main effect of DRK can be decribed as follows. The deformation term 
increases with energy roughly like $\approx ~E^3$ , whereas the "mass term"
in (9) decreases like $\approx ~E^{-1}$ . The ratio between the two terms 
varies like $\approx ~E^4$ and, above some energy depending on the parameters 
involved, the deformation becomes dominant as compared to the mass term. 
Very-high energy kinematics in the laboratory rest frame
is, basically, dominated by longitudinal
momentum: as everything else becomes "small", and longitudinal momentum 
has to be exactly conserved, the real kinematical balances occur entirely
between
"small" terms. Therefore, a "small" violation of the relativity principle can
potentially play a crucial role in these balances.
If $c$ is universal, $\alpha $ must be positive
to avoid the spontaneous decay of UHE (ultra-high energy) particles
(unless the effect is not observable below $3.10^{20}~eV$). The universality 
of $\alpha $ up to small
corrections is imposed by the requirement that elementary particles
be able to reach very high energies (again, if 
the effect is to be obervable below
$3.10^{20}~eV$). Otherwise, particles with smaller positive values of
$\alpha$ would decay into those with larger $\alpha$ (Gonzalez-Mestres,
1997e) and the effect would manifest itself in high-energy cosmic-ray events.      

~
\subsection{Alternative models}
\label{am.sec}

~

The above model is not the only possible way to deform relativistic 
kinematics. Alternatives are:

{\it Mixing with superluminal sectors (MSLS).} 
A form of DRK was predicted in our papers since 1995, 
where we attributed (Gonzalez-Mestres, 1995) Lorentz symmetry violation
to a very high-energy, very low-distance phenomenon which would modify
the propagators of "ordinary" particles (those with critical speed in
vacuum equal to $c$ , the speed of light): the dynamics driving LSV was 
expected to be generated at Planck scale or at some other fundamental
length scale. The
energy-dependence of LSV was claimed to be the explanation to the 
apparent validity of the Poincar\'e relativity principle, as inferred
from low-energy tests.
We also pointed out (see, 
e.g. Gonzalez-Mestres, 1996) that ultra-high-energy cosmic
rays would be a natural experimental framework
to explore possible Lorentz symmetry
violation phenomena. A LSV scenario suggested in all these papers was
mixing between "ordinary" and superluminal particles directly deforming
propagators (see also   
Gonzalez-Mestres, 1997d where energy-dependent mixing parameters
were explicitly used, preserving Lorentz symmetry 
in the $k~\rightarrow 0$ limit). 
Using such models, counterexamples to the claims made in (Coleman and Glashow,
1997) were presented, based on the energy-dependence of LSV effective 
parameters. Through the parametrizations thus 
considered, it was also pointed out that,
besides a fundamental length scale, masses of heavy superluminal particles
can play a significant role in killing low-energy LSV.  

The basic idea of our superluminal particle 
({\it superbradyons}, see e.g. Gonzalez-Mestres, 1997g)
model was that several sectors
of matter are generated at the fundamental length scale(s), each sector
possibly satisfying a "sectorial" Lorentz invariance vith a "sectorial"
critical speed in vacuum ($c_i$ for the $i$-th superluminal sector).
Superbradyons would have positive mass and energy, and satisfy
sectorial motion equations (e.g. Klein-Gordon) with critical speed
$c_i$ : thay are 
not tachyons. 
Dynamical mixing between two sectors would break both Lorentz invariances, 
and
mixing with superluminal sectors (MSLS) would be enough to produce a consistent
DRK for "ordinary" matter. But, if LSV is generated in this way, 
we also 
expect more conventional deformations of particle propagators to occur,
other
than direct mixing between different sectors of matter.  

{\it Linear deformation.} 
When building (1997) the DRK approach given by (7)-(9), where
the effective deformation parameter depends quadratically on energy 
(quadratically deformed relativistic kinematics, QDRK),
we were also naturally led to consider models where this dependence is linear
(linearly deformed relativistic kinematics, LDRK), i.e. where $e~(k~a)$ is a 
function of $k~a$ and, for $k~a~\ll ~1$ : 
\begin{eqnarray}
e~(k~a) ~ \simeq ~ [(k~a)^2~-~\beta ~(k~a)^3~
+~(2\pi ~a)^2~h^{-2}~m^2~c^2]^{1/2}
\end{eqnarray}
\noindent
$\beta $ being a model-dependent constant. 
For momentum $p~\gg ~mc$ :
\begin{eqnarray}
E ~ \simeq ~ p~c~+~m^2~c^3~(2~p)^{-1}~
-~p~c~\beta ~(k~a)/2~~~~~
\end{eqnarray}
\noindent
the deformation $\Delta~E~=~-~p~c~\beta ~(k~a)/2$ being now driven by an 
effective parameter proportional to momentum,
$\Gamma~(k)~=~\Gamma^l_0~k$ where $\Gamma^l_0 ~=~-~\beta ~a/2$ . 

QDRK naturally emerges when a fundamental length scale is introduced to
deform the Klein-Gordon equations. It is typical, for instance, of phonons 
in condensed-matter physics. As recently pointed out (Ellis et al., 
1999a and b) using a class of string models, LDRK can be generated by 
introducing a background 
gravitational field in the propagation equations of free particles. 
In the first case, the Planck scale is an internal parameter of the basic
wave equations generating the "elementary"
particles as vacuum excitations. In the second
case, it manifests itself only as a parameter of the background gravitational
field, similar to a refraction phenomenon. By discriminating between the two 
parametrizations, or excluding both 
approaches,
feasible experiments at available energies can potentially provide very 
valuable information on fundamental Planck-scale physics. Our choice in
1997 was to concentrate on the QDRK model and disregard LDRK, for 
phenomenological 
reasons which seem to remain valid if the new physics is expected to be 
generated not too far from Planck scale.

If existing bounds on LSV from
nuclear magnetic resonance experiments are to be intepreted as setting a
bound of $\approx 10^{-21}$ on relative LSV at the momentum scale 
$p~\sim ~100~MeV$ , this implies $\beta ~a~<~10^{-34}~cm$ . However, as it
will be explained later, it turns out that LDRK can lead to many 
inconsistencies with cosmic-ray experiments unless $\beta ~a$ is much 
smaller. 
Concepts and 
formulae presented in our previous papers for QDRK (see next section)
can be readily extended to LDRK, using similar techniques. In 
particular:

- the linear deformation term $-~p~c~\beta ~(k~a)/2$ and the mass term 
$m^2~c^3~(2~p)^{-1}$ become of the same order at the energy scale
$E_{trans}~
\approx ~\pi ^{-1/3}~ h^{1/3}~(2~\beta )^{-1/3}~a^{-1/3}~m^{2/3}~c^{5/3}$ ;

- the linear deformation term and the target energy $E_T$
become of the same 
order at the energy scale
$E_{lim} ~\approx ~(2~\pi )^{-1/2}~(E_T~a^{-1} \beta ^{-1}~h~c)^{1/2}$ .

- if the same philosophy as for QDRK is to be 
followed, $c$ and $\beta$ would be universal for all "elementary" particles
including light hadrons, whereas for larger objects $\beta \propto m^{-1}$ ,
the nucleon mass setting the scale.
 
{\it Modifications of QDRK.} Our conjecture that $\alpha$ has the same
value for light hadrons as for the photon and leptons derives from the result
(Gonzalez-Mestres, 1997f)
that the elementary soliton solution on a one-dimesional space
lattice obeys the
same deformed 
kinematics as plane waves on the same lattice and with the same 
dalembertian operator (discretized in space), if the soliton size scale is
basically the quantum inverse of its mass scale. As the highest-energy observed
cosmic-ray events seem to be hadronic and not electromagnetic, the conjecture
seems sensible on practical grounds. 
But it can also be argued
that quarks are the real elementary particles 
and that, to be consistent with quark
propagation, the value of $\alpha$ should be divided by a factor
$\approx 4$ for mesons and $\approx 9$ for baryons. 
Actually, the parton
picture seems impossible to implement when the deformation becomes 
important, as in the conventional parton model the constituents can 
carry arbitrary fractions of energy and momentum at very high energy
and the deformation energy depends crucially on these fractions
(Gonzalez-Mestres, 1997f): we therefore expect the failure of any parton 
model at these energies. 

A phenomenological discussion of the latter hypothesis
is nevertheless worth attempting: for
instance, if alpha is to be divided by a factor of 9
for the proton and 4 for the pion, 
and the highest-energy cosmic-ray events are due to protons,
the requirement that the spontaneous decay $p~\rightarrow~p~+~\gamma$
does not occur would then imply $\alpha ~a^2~<~2.10^{-73}~cm^2$ . 
This would exclude
LSV with strong coupling at the Planck scale.
A similar kinematical analysis seems to hold for nuclei
unless dynamics prevents the decay (see next section). 
With 
$\alpha ~a^2~=~2.10^{-73}~cm^2$ , a proton with $E
~\approx~10^{21}~eV$ could also emit pions and, above $E
~\approx~10^{20}~eV$ ,
pion lifetimes would become much shorter than predicted by special relativity
and charged pions can emit photons. 

~

\section{QDRK and Ultra-High Energy Cosmic-Ray (UHECR) Physics}
\label{uhcr.sec}

~

If Lorentz symmetry is broken at Planck scale or at some other
fundamental length scale, the effects of LSV may be accessible to experiments
well below this
energy: in particular, they can produce detectable phenomena at the highest
observed cosmic ray energies. DRK
(Gonzalez-Mestres 1997a, 1997b, 1997c, 1997h and 1998a) plays a crucial
role. Taking the quadratic deformation (QDRK) version of DRK, it is found
that, at energies above
$E_{trans}~
\approx ~\pi ^{-1/2}~ h^{1/2}~(2~\alpha )^{-1/4}~a^{-1/2}~m^{1/2}~c^{3/2}$,
the very small deformation $\Delta ~E$
dominates over
the mass term $m^2~c^3~(2~p)^{-1}$ in (3) and modifies all
kinematical balances: 
physics gets thus closer to Planck scale than
to electroweak scale (this is actually the case in a logarithmic plot of
energy scales)
and UHECR become an efficient probe of Planck-scale physics. 
Because of the negative value of $\Delta ~E$ , it costs
more and more energy, as energy increases above $E_{trans}$,
to split the incoming logitudinal momentum in the laboratory rest frame.
As the ratio $m^2~c^3~(2~p~\Delta ~E)^{-1}$ varies like $\sim ~E^{-4}$ ,
the transition at $E_{trans}$ is very sharp.

With such a LSV pattern, we also inferred (Gonzalez-Mestres, 1997f)
from a toy soliton
model that 
the parton picture (in any version), as well as standard
relativistic 
formulae for Lorentz contraction and time dilation, are expected to fail
above this energy (Gonzalez-Mestres, 1997b and 1997f) which corresponds to $E
~\approx~10^{20}~eV$
for $m$ = proton mass and
$\alpha ~a^2~\approx ~10^{-72}~cm^2$ (f.i. $\alpha
~\approx ~10^{-6}$ and $a$ = Planck
length), and to $E~\approx ~10^{18}~eV$ for
$m$ = pion mass and $\alpha ~a^2~\approx ~10^{-67}~cm^2$
(f.i. $\alpha ~\approx ~0.1$ and $a$ = Planck length). Such effects are
in principle detectable. A phenomenological
study of the implications 
of DRK allowed to draw several important conclusions for UHECR 
experiments (Gonzalez-Mestres, 1997-99).

~

\subsection{Our previous predictions with QDRK}
\label{prev.sec}

~

Assuming that the earth moves slowly with
respect to the absolute rest frame
(the "vacuum rest frame"), so that the approximation (9) remains valid
in the laboratory rest frame, QDRK can 
lead to observable phenomena
in future experiments devoted to the highest-energy cosmic rays:

a) For $\alpha ~a^2~>~10^{-72}~cm^2$ , 
assuming universal values of $\alpha $ and $c$ ,
there is no Greisen-Zatsepin-Kuzmin (GZK)
cutoff (Greisen 1966; Zatsepin \& Kuzmin, 1966)
for the particles under
consideration. Due to the new kinematics, 
interactions with cosmic microwave background (CMB)
photons are strongly inhibited or
forbidden, and ultra-high energy cosmic rays (e.g. protons)
from anywhere in the presently observable Universe
can reach the earth (Gonzalez-Mestres, 1997a and 1997c). In particular,
for an incoming UHE nucleon hitting a CMB photon, the $\Delta $ resonance 
can no longer be formed due to the deformation term. Proton deceleration
in astrophysical objects (e.g. gamma-ray bursters) can be inhibited in a
similar way.

b) With the same hypothesis,
unstable particles with at
least two stable particles in the final states
of all their decay channels become stable at very
high energy. Above $E_{trans}$, the lifetimes of all
unstable particles (e.g. the $\pi ^0$ in
cascades) become much longer than predicted
by relativistic kinematics (Gonzalez-Mestres, 1997a, 1997b and 1997c).
Then, for instance,
the neutron or even the $\Delta ^{++}$ can be candidates for the
primaries of the highest-energy cosmic ray
events. If $c$ and $\alpha $ are not exactly universal, 
many different scenarios can happen concerning the stability of 
ultra-high-energy particles
(Gonzalez-Mestres, 1997a, 1997b and 1997c).

c) In astrophysical processes at very
high energy,
similar mechanisms can inhibit radiation under
external forces (e.g. synchrotron-like, where the interactions occur with
virtual photons), GZK-like cutoffs, decays,
photodisintegration of nuclei, momentum loss trough
collisions (e.g. with a photon wind in reverse shocks), 
production of lower-energy secondaries...
potentially contributing to solve all basic problems
raised by the highest-energy cosmic rays (Gonzalez-Mestres, 1997e),
including acceleration mechanisms.

d) With the same hypothesis, the allowed final-state
phase space of two-body collisions is strongly 
reduced at very high energy, leading 
(Gonzalez-Mestres, 1997e) to a 
sharp fall of partial and total cross-sections
for incoming cosmic ray energies above
$E_{lim} ~\approx ~(2~\pi )^{-2/3}~(E_T~a^{-2}~ \alpha ^{-1}~h^2~c^2)^{1/3}$,
where $E_T$ is the energy of the target. 
As a consequence, and with the
previous figures for Lorentz symmetry violation parameters, above some
energy $E_{lim}$ between 10$^{22}$ and $10^{24}$ $eV$ a cosmic
ray will not deposit most of its energy in the atmosphere
and can possibly fake an exotic event with much less energy
(Gonzalez-Mestres, 1997e).

e) Actually, requiring simultaneously the absence of GZK cutoff in the region 
$E~\approx ~
10^{20}~eV$ , and that cosmic rays with
energies below $\approx ~3.10^{20}~eV$ deposit most of their energy in the
atmosphere, leads in the DRK scenario to the constraint:
$10^{-72}~cm^2~<~\alpha ~a^2~<~
10^{-61}~cm^2$~, equivalent to $10^{-20}~<~\alpha ~<~10^{-9}$ for
$a~\approx 10^{-26}~cm$~ ($\approx~10^{21}~GeV$ scale). 
Remarkably enough, assuming full-strength
LSV forces $a$ to be in the range $10^{-36}~cm~<~a~<~
10^{-30}~cm$~. But a $\approx 10^{-6}$ LSV at Planck scale
can still explain the data. Thus, the simplest
version of QDRK naturally fits, on
phenomenological grounds, with the expected potential
role of Planck scale in
generating the standard "elementary" particles and opening 
the door to new physics.  

f) Effects a) to e) are obtained applying DRK to single particles and 
collisions. If further dynamical
anomalies are added (failure, at very small distance scales,
of the parton model and of the
standard relativistic formulae for Lorentz contraction and time
dilation...), we can expect
much stronger effects in the early cascade development profiles
of cosmic-ray events (Gonzalez-Mestres, 1997b, 1997f and 1998a).
Detailed phenomenology and
data analysis in next-generation experiments may uncover 
spectacular new physics and provide a powerful microscope directly
focused on the fundamental length (Planck?) scale.

g) Cosmic superluminal particles would produce atypical events
with very small total momentum (due to the high $E/p$
ratio), isotropic or involving several
jets 
(Gonzalez-Mestres, 1996, 1997b, 1997d, 1997 and 1998b). In the atmosphere (f.i. 
AUGER or satellite-based experiments),
such events would generate exceptional cascade development profiles 
and muon spectra,
as will be discussed in a forthcoming paper.

~

\noindent
***************************************************************************************

It should be noticed that our description of all these phenomena and, in
particular, of points a) and b) , on the grounds of DRK was prior to
any similar claim by Coleman and Glashow from $TH\epsilon \mu $-like models.

\noindent
***************************************************************************************

~

It follows from b) and f) 
that early cascade development is a crucial point for 
possible
tests of special relativity in UHCR cosmic-ray events. Unfortunately, this part of the
interactions induced by the incoming cosmic ray is not easily detectable: 
it occurs early in the atmosphere, and it takes
a few collisions before energy is degraded into a large enough number of 
particles to produce an observable fluorescence signal. 
Data on cascade development start well below the first 
interaction of
the primary with the atmosphere whereas we expect LSV, if present, 
to manifest itself
only in the first few collisions. But the effect will propagate to later
multiparticle
production and be observable: for instance,
if the $\pi ^0$ does not decay at very high energy,
we expect a smaller electromagetic component developing later and more muons 
produced. 
A serious drawback is the present ambiguousness of phenomenological air 
shower models,
but combined data from AUGER and satellite-based experiments should 
help to
clarify this situation. A recent fit to the UHECR spectrum 
with a model close to 
QDRK, reproducing the absence of GZK cutoff, can 
be found in (Chechin \& Vavilov, 1999).

~

\subsection{On cosmic-ray composition}
\label{comp.sec}

~

Cosmic-ray composition above $10^{17}~eV$ is a crucial question. 
Protons are often preferred as candidates to UHECR events (Bird et al.,
1993), but detailed analysis are not yet conclusive (f.i. AGASA 
Collaboration, 1999)
and there are claims in favor of light nuclei up to 2.10$^{19}~eV$ 
(f.i. Wolfendale \& Wibig, 1999). It is commonly agreed that the UHECR
composition becomes lighter above $5.10^{17}~eV$ , as energy increases,
and tends to be protons at the highest observed energies. It may be
interesting to compare this phenomenon with QDRK predictions, keeping in 
mind the proposed $\alpha \propto m^{-2}$ law. This law naturally
allows, kinematically, spontaneous
$N~\rightarrow ~N~+~\gamma$ decays ($N$ = nucleus) at very high energy. 
If $\alpha ~a^2~\approx ~10^{-67}~cm^2$ , the threshold for
spontaneous gamma emission would be $\approx ~3.10^{19}~eV$ for $Fe$  
and $\approx ~10^{19}~eV$ for $He^4$ . 

It should be noticed, however, that although spontaneous gamma emission
will be kinematically allowed for large objects because 
of the $\alpha \propto m^{-2}$ law, it will become more and
more unnatural dynamically as their size increases. Such an
outgoing photon 
would carry an energy
much larger than that of any incoming nucleon and its wavelength
would be smaller, by orders of magnitude, 
than the size of the composite
object.  

~ 

\subsection{Neutrinos from gamma-ray bursts}
\label{nu.sec}

~

The basic idea (Waxman and Bahcall, 1999) is that, in reverse shocks,
protons are accelerated to energies $\sim ~10^{20}~eV$ and collide with
ambient photons producing, in the kinematical region where the center of
mass energy corresponds to the $\Delta ^+$ resonance, charged pions with about
20$\%$ of the initial proton energy. These pions subsequently decay into muons
and muon neutrinos ($\pi^+~\rightarrow~\mu^+~+~\nu_\mu $),
and the muons decay later into electrons, electron neutrinos
and muon antineutrinos ($\mu^+~\rightarrow ~e^+~+~\nu_e ~+~\overline\nu_\mu $) . 
It should be noticed that, beacuse of Lorentz dilation,
the lifetime for the first decay at pion energy $\sim 2.10^{19}~eV$ is
$\sim 3000 s$ whereas the muon lifetime at $\sim 10^{19}~eV$ is $\sim 10^5~s$ .
These time scales are already 
longer than those of gamma-ray bursts. But the situation,
for UHE neutrino production, may be considerably worsened by LSV. For 
$\alpha ~a^2~>~10^{-72}~cm^2$ , the lifetime of a $\sim ~10^{20}~eV$ 
$\Delta ^+$ is modified
by QDRK and gets much longer. If $\alpha ~a^2~>~10^{-70}~cm^2$ , the resonance 
becomes stable at the same energy with respect to the 
$n~e^+~\nu _{\mu }~\overline\nu_ {\mu}~\nu _e$ decay channel. Also, as
previoulsy stated, much higher photon energies are required to form a
$\Delta $ resonance. 
In both cases, pion photoproduction is strongly inhibited and the calculation
presented in (Waxman and Bahcall, 1999) is to be modified leading to a lower
neutrino flux. 
Similarly, the lifetimes of UHE charged pions
and muons become much longer and lower the neutrino flux further.  
At the same time, QDRK
may inhibit UHE proton syncrotron radiation and allow the proton to be
accelerated to energies above $\sim ~10^{20}~eV$ .

As DRK makes velocity, $v~=~dE/dp$ , energy-dependent (Gonzalez-Mestres, 1997a
and 1997d), 
the arrival time on
earth of particles produced in a single burst 
is expected to dependent on the particle 
energy. 
But, contrary to the claim presented in (Ellis et al., 1999a), it follows from the
above considerations that observing
UHE neutrinos from GRB bursts for QDRK with $\alpha ~a^2~>~10^{-72}~cm^2$  
will most likely be much more difficult (if at all possible) than  
naively expected from the Waxman-Bahcall model.  

~

\section{LDRK, gamma-ray bursts and TeV physics}
\label{grb.sec}

~

Considering vacuum as a medium similar to an electromagnetic plasma,
it was suggested in (Amelino-Camelia et al., 1998) that quantum-gravitational 
fluctuations may lead to a correction, linear in energy, to the velocity of 
light. This is equivalent to a LDRK that, for $k~a~\ll ~1$ , 
can be parameterized as:

\begin{eqnarray}
E ~\simeq ~p~c~-~p~c~\beta ~(k~a)/2~=~p~c~-~p^2~M^{-1}
\end{eqnarray}
 
\noindent
where $M$ is an effective mass scale. Possible tests of this model through
gamma-ray bursts, measuring the delays in the arrival time of photons of
different energies, have been considered in (Norris et al., 1999) having in
mind the Gamma-ray Large Area Space Telescope (GLAST) and in
(Ellis et al., 1999b) with a more general scope. Biller et al. (1998) claim
a lower bound on $M$ slightly above $\approx 10^{16}~GeV$ . 

However, from the same considerations already developed in our 1997-99
papers taking QDRK as an exemple, stringent bounds on LDRK can be 
derived. 
Assume that LDRK applies only to photons, and not to charged particles, so
that at high energy we can write for a charged particle, $ch$ ,
the dispersion 
relation: 

\begin{eqnarray}
E_{ch} & \simeq & p_{ch}~c~+~m_{ch}^2~c^3~(2~p_{ch})^{-1}
\end{eqnarray}
\noindent
where the $ch$ subscript stands for the charged particle under consideration.
Then, it can be readily checked that the decay $ch~\rightarrow ~ch~+~\gamma$ 
would be allowed for $p$ above $\simeq ~(2~m_{ch}^2~M~c^3)^{1/3}$ , i.e: 

- for an electron, above $E ~\approx ~2~TeV$ if $M~=10^{16}~GeV$ 
and $\approx ~20~TeV$ if $M~=10^{19}~GeV$ 

- for a muon or charged pion, above $E ~\approx ~80~TeV$ if $M~=10^{16}~GeV$
and $\approx ~800~TeV$ if $M~=10^{19}~GeV$

- for a proton, above $E ~\approx ~240~TeV$ if $M~=10^{16}~GeV$
and $\approx ~2.4~PeV$ if $M~=10^{19}~GeV$

- for a $\tau $ lepton, above $E ~\approx ~400~TeV$ if $M~=10^{16}~GeV$
and $\approx ~4~PeV$ if $M~=10^{19}~GeV$

\noindent
so that none of these particles would be oberved above such energies, 
apart from very short paths.
Such decays seem to be in obvious contradiction with cosmic ray data, but
avoiding them forces the charged particles to have the same kind of 
propagators as the photon, with the same effective value of $M$ up to small
differences. Similar conditions are readily derived for all 
"elementary" particles, leading for all of them, up to small devations, 
to a LDRK given by the universal dispersion relation:

\begin{eqnarray}
E ~\simeq ~p~c~+~m^2~c^3~(2~p)^{-1}~-~p^2~M^{-1}
\end{eqnarray}

For instance, $\pi^0$ production would otherwise be 
inhibited. But if, as it seems compulsory, the $\pi^0$ kinematics   
follows a similar law, then the decay time for $\pi^0 ~\rightarrow ~\gamma
~\gamma $ will become much longer than predicted by special relativity at
energies above  $\approx ~50~TeV$ if $M~=10^{16}~GeV$
and $\approx ~500~TeV$ if $M~=10^{19}~GeV$ . Again, this seems to 
be in contradiction
with cosmic-ray data. Requiring that the $\pi^0$ lifetime agrees
with special relativity at $E ~\approx ~10^{17}~eV$ would force $M$
to be above $\approx ~10^{26}~GeV$ , far away from the values to be
tested at GLAST. Another bound is obtained from the
condition that there are $3.10^{20}~eV$ cosmic-ray events. Setting
$E_{lim}$ to this value, and taking oxygen to be the 
target, yields $M~\approx ~ 
3.10^{21}GeV$ . In view of these bounds, it
appears very difficult to make LDRK , with $M$
reasonably close to Planck scale, compatible with experimental data.
It therefore seems necessary to reconsider the models to be tested at GLAST.
~

\section{Conclusions and comments}
\label{c.sec}

~

If, as conjectured by Poincar\'e, 
special relativity is a symmetry of dynamical origin,
Lorentz symmetry violation at very high energy  
would not be unnatural. But checking it appears to be
a difficult task. If LSV is generated at some fundamental length (e.g.
Planck) scale, we expect it to be driven by energy-dependent parameters
becoming very small in the low-energy region where impressive tests of
the validity of the relativity principle have been performed (Lamoreaux et al.,
1986; Hills \& Hall, 1990). It therefore seems natural,
contrary to the $TH\epsilon \mu $ model, to preserve Lorentz
symmetry as a low-energy limit. 
In deformed Lorentz symmetry models, leading in particular 
to various versions of deformed relativistic 
kinematics, the deformation disappears in the limit
$k~\rightarrow 0$ . DRK is far from being unique, as it can be generated in 
many different ways, and its predictions are
strongly model-dependent. Unlike a previous attempt (Kirzhnits \& Chechin,
1972), we have pursued the idea that: a) the fundamental length scale is 
at the origin of LSV and should be taken seriously in all respects; 
b) physics can vary smoothly down to this scale where a new dynamics manifests
itself; c)
LSV is not related to any other fundamental symmetry generalizing special
relativity. It then turns out that, at energies well below the fundamental
length scale, a very small LSV can generate observable
leading-order effects and even allow to discriminate between models of vacuum
at Planck scale.  

Ambitious prospects, based on LDRK (linearly deformed relativistic kinematics,
obtained from Planck-scale "vacuum recoil" models), to measure a possible
systematic energy-dependence in time delays
from gamma-ray bursts are confronted to apparent incompatibilities between
cosmic-ray data and the orders 
of magnitude of LDRK parameters considered. The basic reason is that
LSV, if realized in this way, would manifest itself in many other phenomena
already accessible to experiment (cosmic rays in the TeV - PeV range). However,
the fact that the study 
of LDRK has led to consider such comparatively low
energies pleads in favour of precision tests of special relativity at the 
highest-energy accelerators (LHC, VLHC and beyond). 
Although the existence of observable
effects of LDRK at the gamma-ray burst photon energies seems unlikely,
systematic tests of special relativity at energies between 1 TeV and 1 PeV
are missing and should be performed.

On phenomenological grounds, QDRK (quadratically deformed relativistic
kinematics, obtained when a fundamental length scale is introduced 
in the dynamics generating "elementary" particles) seems
to be the most performant model and naturally fits with possible new
Planck-scale physics. 
UHE (ultra-high energy) CR (cosmic-ray) physics appears to be the safest
laboratory to test LSV, allowing for unconventional phenomena in 
early cascade development. If Lorentz symmetry is violated,
the study of UHECR events may be a powerful  
tool to get direct information
on fundamental physics at Planck scale. 
At the highest oberved cosmic-ray energies, the effect of LSV at Planck scale 
can already lead to spectacular signatures: absence of the GZK cutoff;
drastic modifications of lifetimes, as well as of
total and partial cross-sections;
failure of any parton picture as well as of standard formulae for
time dilation and Lorentz contraction... 
But further
work is required to clearly translate the physical
signatures into measurable data
(cascade development profile, muons, electromagnetic yield...). 
Models of air shower formation should be improved in order 
to remove many interpretation
uncertainties. 
A basic difficulty is that, although the primary interaction occurs quite 
early in the atmosphere, a detectable fluorescence yield is emitted only
after a few interactions, when many charged particles have been produced and 
the atmosphere is denser.
Combined information from future experiments, such as AUGER and satellite-based 
measurements, will hopefully make this
task easier. 

The existence of superluminal particles ({\it superbradyons})
is not excluded if LSV occurs 
at Planck scale: they may even be the ultimate constituents
of matter. This subject
will be discussed at length elsewhere.

%
%
%
\vspace{1ex}
\begin{center}
{\Large\bf References}
\end{center}
%
\noindent
AGASA Collaboration, 1999, Proceedings of the 26$^{th}$ International Cosmic
Ray Conference (ICRC 1999), Salt Lake City August 1999, 
paper HE 2.3.02.\\ 
Amelino-Camelia, G., Ellis, J., Mavromatos, N.E., Nanopoulos, D.V. 
\& Sarkar, S.,
Nature 393, 319 (1998). \\
Biller, S.D., Breslin, A.C., Buckley, J., (Biller et al.), 
1998, paper gr-qc/9810044 .\\
Bird, D.J., Corbato, S.C., Dhai, H.Y. {\it et al.}
(Bird et al.), 1993, Proceedings of the 23$^{th}$ International Cosmic 
Ray Conference (ICRC 1993), Calgary 1993, Vol. 2, p. 38.\\
Chechin, V.A., \& Vavilov, Yu.N., ICRC 1999 Proceedings, paper HE 2.3.07.\\
Coleman, S., \& Glashow, S., 1997, Phys. Lett. B 405, 249 and subsequent 
(1998)
papers
in LANL (Los Alamos) electronic archive as well as in the TAUP 97 
Proceedings.\\
Einstein, A., 1905, "Zur Elecktrodynamik bewegter K\"{o}rper",
Annalen der Physik 17, 891.\\
Einstein, A., 1921, "Geometrie und Erfahrung",
{\it Preus. Akad. der Wissench.,
Sitzungsberichte}, part {\bf I}, p. 123. \\
Ellis, J., Mavromatos, N.E., Nanopoulos, D.V. \& Volkov, G. (Ellis et al.)
1999a, paper
gr-qc/9911055.\\
Ellis, J., Farakos, K., Mavromatos, N., Mitsou, V.A. \& Nanopoulos, D.V. 
(Ellis et al.), paper astro-ph/9907340.\\   
Feynman, R.P., Leighton, R.B., \& Sands, M., 1964 
"The Feynman Lectures on Physics", Addison-Wesley.\\
Greisen, K., 1966, {\it Phys. Rev. Lett.} 16 , 748.\\ 
Gonzalez-Mestres, L., 1995, paper astro-ph/9505117 of LANL electronic archive,
Proceedings of the Moriond Workshop on "Dark Matter, Cosmology, Clocks
and Fundamental Laws", January 1995, Ed. Fronti\`eres.\\
Gonzalez-Mestres, L., 1996, paper hep-ph/9610474 of LANL 
electronic archive and references therein.\\
Gonzalez-Mestres, L., 1997a, paper
physics/9704017 of LANL archive.\\
Gonzalez-Mestres, L., 1997b, Proc. of the International
Conference on Relativistic Physics and some of its Applications, Athens June
25-28 1997, Apeiron, p. 319.\\
Gonzalez-Mestres, L., 1997c, Proc. 25th ICRC (Durban, 1997), Vol. 6, p. 113.\\
Gonzalez-Mestres, L., 1997d, papers physics/9702026
and physics/9703020 of LANL archive.\\
Gonzalez-Mestres, L., 1997e, papers physics/9706022 and
physics/9706032 of LANL archive.\\
Gonzalez-Mestres, L., 1997f, paper nucl-th/9708028 of LANL archive.\\
Gonzalez-Mestres, L., 1997g, Proc. 25th ICRC (Durban, 1997), Vol. 6, p. 109.\\
Gonzalez-Mestres, L., 1997h, 
talk given at the International
Workshop on Topics on Astroparticle and Underground Physics (TAUP 97),
paper physics/9712005 of LANL archive.\\
Gonzalez-Mestres, L., 1998a, talk given at the "Workshop on Observing 
Giant Cosmic Ray Air Showers From $>$ 10$^{20}$ Particles From Space", College
Park, November 1997, AIP
Conference Proceedings 433, p. 148.\\
Gonzalez-Mestres, L., 1998b, same Proceedings, p. 418.\\ 
Gonzalez-Mestres, L., 1998c, Proc. of COSMO 97, Ambleside September 1997,
World Scientific, p. 568.\\
Gonzalez-Mestres, L., 1999a,  
ICRC 1999 Proceedings, paper HE. 1.3.16.\\
Gonzalez-Mestres, L., 1999b, ICRC 1999 Proceedings, paper OG 3.1.10.\\
Hills, D. \& Hall, J.L., 1990, {\it Phys. Rev. Lett.} 64 , 1697.\\
Kirzhnits, D.A., and Chechin, V.A., 1972,
{\it Soviet Journal of Nuclear Physics},
15 , 585.\\
Lamoreaux, S.K., Jacobs, J.P., Heckel, B.R., Raab, F.J. \& Forston, E.N.,
1986, {\it Phys. Rev. Lett.} 57 , 3125.\\
Logunov, A.A., 1995, "On the articles by Henri Poincar\'e on the
dynamics of the electron", Ed. JINR, Dubna.\\
Logunov, A.A., 1997,  
"Relativistic theory of gravity and the
Mach principle", Ed. JINR, Dubna.\\
Miller, A.I. 1996, Why did Poincar\'e not formulate Special
Relativity in 1905?", in "Henri Poincar\'e, Science and Philosophy", Ed.
Akademie Verlag, Berlin, and Albert Blanchard, Paris.\\
Norris, J.P., Bricell, J.T., Marani, G.F. \& Scargle, J.D.,
1999, ICRC 1999 Proceedings, paper OG 2.3.11.\\
Paty, M., 1996, "Poincar\'e et le principe de relativit\'e",
same reference.\\
Peebles, P.J.E., 1993, "Principles of Physical Cosmology", 
Princeton University Press.
Poincar\'e, H., 1895, "A propos de la th\'eorie de M. Larmor",
L'Eclairage \'electrique, Vol. 5,  5.\\
Poincar\'e, H., 1901, "Electricit\'e et Optique: La lumi\`ere
et les th\'eories \'electriques", Ed. Gauthier-Villars, Paris.\\
Poincar\'e, H., 1905, Sur la dynamique de l'\'electron", Comptes
Rendus Acad. Sci. Vol. 140,
p. 1504, June 5.\\
Waxman, E. \& Bahcall, J., 1999, paper hep-ph/9909286.\\
A.W. Wolfendale \& T. Wibig, 1999, ICRC 1999 Proceedings, paper OG 1.3.01.\\
Will, C., 1993, "Theory and Experiment in
Gravitational Physics", Cambridge University Press.\\
Zatsepin, G.T. and
Kuzmin, V.A., {\it Pisma Zh. Eksp. Teor. Fiz.} 4 , 114.
\end{document}